\documentclass[aps,prl,twocolumn,superscriptaddress,longbibliography,floatfix]{revtex4-2}

\usepackage{amssymb}
\usepackage{amsmath}
\usepackage[svgnames,dvipsnames,x11names]{xcolor}
\usepackage{hyperref}
\hypersetup{
colorlinks=true,
citecolor= Maroon,
linkcolor= MidnightBlue,
urlcolor= Maroon,
pdfauthor={},
pdftitle={},
pdfsubject={}
}

\newcommand{\OGr}{\mathrm{OGr}}
\newcommand{\Res}{\mathop{\mathrm{Res}}}

\newcommand{\GL}{\mathrm{GL}}

\newcommand{\Ycal}{\mathcal{Y}}
\newcommand{\Del}{\Delta}

\newcommand{\ls}{\ell_s}
\newcommand{\lt}{\ell_t}
\newcommand{\lu}{\ell_u}
\newcommand{\lbs}{\bar\ell_s}
\newcommand{\lbt}{\bar\ell_t}
\newcommand{\lbu}{\bar\ell_u}
\newcommand{\es}{e_s}
\newcommand{\ebs}{\bar e_s}
\newcommand{\et}{e_t}
\newcommand{\ebt}{\bar e_t}
\newcommand{\eu}{e_u}
\newcommand{\ebu}{\bar e_u}
\newcommand{\tbs}{\bar\tau_s}
\newcommand{\tbt}{\bar\tau_t}

\begin{document}
\title{The Vasiliev Grassmannian}
\author{Shounak De}
\email{sde25@sas.upenn.edu}
\affiliation{Center for Particle Cosmology, Department of Physics and Astronomy, University of Pennsylvania, Philadelphia, PA 19104, USA}
\affiliation{School of Natural Sciences, Institute for Advanced Study, Princeton, NJ 08540, USA}
\author{Hayden Lee}
\email{haydenhl@sas.upenn.edu}
\affiliation{Center for Particle Cosmology, Department of Physics and Astronomy, University of Pennsylvania, Philadelphia, PA 19104, USA}
\begin{abstract}
We express the scalar four-point function of minimal Vasiliev higher-spin gravity in de~Sitter space as an integral over the orthogonal Grassmannian $\OGr(4,8)$.
The full crossing-symmetric Vasiliev Grassmannian correlator is given by $ (S^2+T^2+U^2)/(STU)$, where $S$, $T$, $U$ are the Grassmannian Mandelstam variables.
Remarkably, this has the same form as the field-theory limit of the Veneziano amplitude, despite arising from the opposite, tensionless limit of an infinite massless higher-spin tower.
We verify the formula by evaluating the Grassmannian contour integral and matching it to the momentum-space result, and analyze its singularities and residues directly in Grassmannian~space.
\end{abstract}
\maketitle

\section{Introduction}

One of the deepest lessons of string theory is that infinite towers of resonances can resum into expressions that are simpler than any finite truncation.
The Veneziano amplitude~\cite{Veneziano:1968yb} packages the entire Regge trajectory into a single Euler beta function, and the simplicity of the final answer is far from obvious when one looks at any finite number of resonances in isolation.

In cosmology, no analogous resummation has been found for massive towers or stringy Regge trajectories.
At tree level, the correlators of inflationary perturbations are organized by their energy singularities and factorization properties~\cite{Maldacena:2011nz, Raju:2012zr, Arkani-Hamed:2015bza, Arkani-Hamed:2017fdk, Arkani-Hamed:2018kmz, Baumann:2020dch, Goodhew:2020hob, Jazayeri:2021fvk, Baumann:2021fxj, Baumann:2022jpr,  Salcedo:2022aal}.
This has proved to be a powerful organizing principle, but its standard perturbative implementation builds correlators from individual exchanges rather than treating the full spectrum at once.
A natural question, then, is whether there exists a ``Veneziano correlator'' for cosmology, in which the fully resummed answer is simpler than any of its individual constituents.

Minimal Vasiliev theory in de Sitter (dS) space provides a useful toy model to explore this question.
The bulk theory contains an infinite tower of massless even-spin fields \cite{Vasiliev:1990en, Vasiliev:2003ev}, yet its boundary correlators can be computed exactly.
Its proposed holographic description is the Euclidean Sp$(N)$ model of~\cite{Anninos:2011ui}, which may be viewed as the dS continuation~\cite{Strominger:2001pn} of the higher-spin/vector-model duality in AdS space~\cite{Klebanov:2002ja,Sezgin:2002rt,Giombi:2009wh,Giombi:2012ms}.
In particular, the $Q$-model of~\cite{Anninos:2017eib} formulates the boundary correlators as one-loop polygon integrals at order $1/N$, yielding rational functions of external momenta at all multiplicities~\cite{DeLee2026}.

At four points, the scalar Vasiliev correlator takes the remarkably compact form in momentum space~\cite{Anninos:2017eib}
\begin{equation}
    \psi_4^{\rm Vas}=\frac{(k_1 k_2 {+} k_3 k_4)s + (k_1 k_4 {+} k_2 k_3)t}{st( k_1 k_3 + k_2 k_4 + st)} + \text{cyc.}\,,\label{eq:boxintro}
\end{equation} 
where $k_i \equiv |\boldsymbol k_i|$, $s \equiv |\boldsymbol k_1+\boldsymbol k_2|$, and $t \equiv |\boldsymbol k_2+\boldsymbol k_3|$.
Notably, the correlator is free of total or partial energy poles, and its only nontrivial singular locus is $ k_1k_3 + k_2k_4 + st=0$, which is the unique Landau singularity of the one-loop box in three spatial dimensions~\cite{DeLee2026}.

The orthogonal Grassmannian $\OGr(n,2n)$, recently introduced in~\cite{Arundine:2026fbr}, offers a new kinematic setting in which to understand this simplicity. 
In this framework, the Grassmannian integrand isolates the intrinsic kinematic data of the correlator in a form that closely resembles its flat-space amplitude counterpart, while the contour integral translates this data into the full momentum-space answer. 
In favorable cases, this separates the simplicity of the underlying object from the complexity generated by the integral.
Given the compactness of~\eqref{eq:boxintro}, it is natural to ask whether the correlator takes an equally simple, or even simpler, form in Grassmannian space, one that makes manifest the underlying infinite-spin exchange.

In this letter, we show that it does. Expressed in the Grassmannian Mandelstam variables $S$, $T$, $U$ defined in~\eqref{eq:STUdef}, the full crossing-symmetric scalar Vasiliev correlator in Grassmannian space is given by
\begin{equation}\label{eq:mainresult}
  \mathcal{A}_4^{\rm Vas} = \frac{S^2 + T^2 + U^2}{STU}\,.
\end{equation}
This manifestly crossing-symmetric function, free of total energy singularities, is simpler than any individual finite-spin exchange, cf.~\eqref{eq:WJ}. As a rational resummation of an infinite higher-spin tower, it may be viewed as a first step toward a cosmological analog of the Veneziano amplitude, as envisioned in~\cite{Arkani-Hamed:2018kmz,Baumann:2022jpr}.

This letter is organized as follows. In~\S\ref{sec:box}, we revisit the Vasiliev four-point function in momentum space and clarify its singularity structure.
In~\S\ref{sec:grassmannian}, we review the essential elements of the cosmological Grassmannian.
In~\S\ref{sec:vasiliev}, we present the Vasiliev Grassmannian correlator and show that the contour integral reproduces the exact momentum-space answer.
In~\S\ref{sec:analytic}, we analyze the analytic properties of the result, and in~\S\ref{sec:veneziano} we discuss the analogy with the Veneziano amplitude.
We conclude in~\S\ref{sec:discussion} with a discussion of open directions.

\section{The Vasiliev box}\label{sec:box}

We begin by rederiving the momentum-space Vasiliev four-point function in a way that sheds new light on its analytic structure. 
In the $Q$-model of~\cite{Anninos:2017eib}, the boundary Hilbert space is built from $2N$ bosonic fields living on the future conformal boundary of dS$_4$, with a Gaussian Hartle-Hawking state. Late-time scalar observables are expectation values of the bilinear operators $B_0(k) \propto \int_q :\!Q_q^\alpha Q_{k-q}^\alpha\!:$ built from the momentum-space fields $Q_k^\alpha$ with $\alpha=1,\dots,2N$, and the connected scalar four-point function is computed by Wick's theorem.
The full connected correlator is a crossing-symmetric sum of three cyclic contributions
\begin{equation}\label{eq:fullcorr}
  \psi_4^{\rm Vas} = \psi_4^{(s,t)} + \psi_4^{(t,u)} + \psi_4^{(u,s)}\,,
\end{equation}
labeled by the three pairs of diagonals $(s,t)$, $(t,u)$, $(u,s)$, where
\begin{equation}\label{eq:diagdef}
  s = |\boldsymbol k_1 {+} \boldsymbol k_2|\,,\quad t = |\boldsymbol k_2 {+} \boldsymbol k_3|\,,\quad u = |\boldsymbol k_1 {+} \boldsymbol k_3|\,.
\end{equation}
Here $\psi_4^{\rm{Vas}}$ denotes the in-in correlator, related to the wavefunction coefficient by a shadow transform on each external leg.
Each contribution takes the form (up to an overall normalization) 
\begin{align}\label{eq:boxmomspace}
 \psi_4^{(s,t)} &= 8k_1k_2k_3k_4  
 \!\int\! \frac{\text{d}^3 l}{(2 \pi)^3} \frac{1}{l^2(\boldsymbol l{+}\boldsymbol{k}_1)^2(\boldsymbol l{+}\boldsymbol k_1{+}\boldsymbol k_2)^2(\boldsymbol l{-}\boldsymbol k_4)^2} \nonumber \\
&= \frac{(k_1 k_2 + k_3 k_4)s + (k_1 k_4 + k_2 k_3)t}{st( k_1 k_3 + k_2 k_4 + st)}\,.
\end{align}
To understand the origin of the singularity at $k_1 k_3 + k_2 k_4 + st=0$, it is instructive to examine the box-to-triangle decomposition.
Denoting the box integral in~\eqref{eq:boxmomspace} by $I_{0123}$, it can be written as~\cite{Melrose:1965kb, DeLee2026}
\begin{equation}
I_{0123}
=
\frac{\Ycal_4\!\left[\begin{smallmatrix}1\\2\end{smallmatrix}\right]}{\Del_4}\,I_{123}
+\frac{\Ycal_4\!\left[\begin{smallmatrix}1\\3\end{smallmatrix}\right]}{\Del_4}\,I_{023}
+\frac{\Ycal_4\!\left[\begin{smallmatrix}1\\4\end{smallmatrix}\right]}{\Del_4}\,I_{013}
+\frac{\Ycal_4\!\left[\begin{smallmatrix}1\\5\end{smallmatrix}\right]}{\Del_4}\,I_{012}\,,
\label{eq:box-sum-triangles}
\end{equation}
where the triangle integrals evaluate to
\begin{align}
I_{123}&=\frac{1}{8k_2k_3t}\,,\quad
I_{023}=\frac{1}{8k_3k_4s}\,,\quad \nonumber \\
I_{013}&=\frac{1}{8k_1 k_4 t}\,,\quad
I_{012}=\frac{1}{8k_1k_2s}\,.
\label{eq:triangle-values}
\end{align}
The cut coefficients $\Ycal_4\left[\begin{smallmatrix}i\\j\end{smallmatrix}\right]$ denote the signed $(i,j)$ minor of the modified Cayley matrix
\begin{equation}
\Ycal_4=
\begin{pmatrix}
0 & 1 & 1 & 1 & 1 \\
1 & 0 & -k_1^2 & -s^2 & -k_4^2 \\
1 & -k_1^2 & 0 & -k_2^2 & -t^2 \\
1 & -s^2 & -k_2^2 & 0 & -k_3^2 \\
1 & -k_4^2 & -t^2 & -k_3^2 & 0
\end{pmatrix},
\label{eq:Y4-box}
\end{equation}
obtained by deleting the $i$th row and $j$th column. The determinant $\Del_4 = \Ycal_4\bigl[\begin{smallmatrix}1\\1\end{smallmatrix}\bigr]$ factorizes into four factors 
\begin{equation}\label{eq:Landaufact}
  \Del_4 = F_{++}F_{+-}F_{-+}F_{--}\,,
\end{equation}
where
\begin{equation}
\begin{aligned}\label{eq:Fpm}
  F_{++} &= k_1k_3 {+} k_2k_4 {+} st\,,\  F_{-+} = -k_1k_3 {+} k_2k_4 {+} st\,,\\
  F_{+-} &= k_1k_3 {-} k_2k_4 {+} st\,, \  F_{--} = -k_1k_3 {-} k_2k_4 {+} st\,.
\end{aligned}    
\end{equation}
When the four triangle contributions in~\eqref{eq:box-sum-triangles} are brought to a common denominator, the numerator assembles into $F_{+-}F_{-+}F_{--}\,\mathcal{N}_4$, where
\begin{equation}\label{eq:N4def}
  \mathcal{N}_4 \equiv (k_1k_2{+}k_3k_4)\,s + (k_1k_4{+}k_2k_3)\,t\,.
\end{equation}
Three of the four Landau branches cancel against $\Del_4$, leaving $F_{++}$ as the sole physical denominator. 
This distinction is most transparent in the Feynman parameter representation. On the branches $F_{+-} = 0$, $F_{-+} = 0$, and $F_{--} = 0$, the Feynman parameters necessarily carry mixed signs and therefore cannot pinch the positive integration simplex. Only $F_{++} = 0$ admits real, positive Feynman parameters supporting a genuine normal threshold~\cite{DeLee2026}. Geometrically, $F_{--} = 0$ is the classical Ptolemy relation for a cyclic quadrilateral inscribed in a circle, and we call its sign-flipped analytic continuation $F_{++} = 0$ the ``Ptolemy singularity''.

\section{The cosmological Grassmannian}\label{sec:grassmannian}

We now briefly review the cosmological Grassmannian introduced in~\cite{Arundine:2026fbr}, focusing on the elements needed for the four-point analysis.
The kinematics of an $n$-point function on the boundary of dS$_4$ is encoded in the $2n \times 2$ spinor matrix $\Lambda = (\lambda_i^\alpha, \bar\lambda_i^\alpha)^T$, where $\lambda, \bar \lambda$ are cosmological spinor helicity variables in $d=3$~\cite{Maldacena:2011nz,Baumann:2020dch}.
Conformal invariance requires correlators to depend on $\Lambda$ only through the combination $C \cdot \Lambda$, where $C$ is an $n \times 2n$ matrix satisfying the orthogonality condition $C \cdot Q \cdot C^T = 0$ with $Q = \bigl(\begin{smallmatrix} 0 & \mathbf{1} \\ \mathbf{1} & 0 \end{smallmatrix}\bigr)$.
The matrix $C$ is defined up to left multiplication by $\GL(n)$, and the resulting equivalence classes parametrize the orthogonal Grassmannian $\OGr(n,2n)$.
Integrating over $C$ produces momentum-space correlators as Grassmannian contour integrals~\cite{Arkani-Hamed:2009ljj, Arundine:2026fbr},
\begin{equation}\label{eq:Grassint}
  \psi_n(\Lambda) = \int \frac{\text{d}^{n\times2n}C}{\GL(n)}\,\delta(C \cdot Q \cdot C^T)\,\delta(C \cdot \Lambda)\,\mathcal{A}_n(C)\,,
\end{equation}
where the integrand $\mathcal{A}_n$ is called the \emph{Grassmannian correlator} and depends on $C$ only through its $n \times n$ minors.
The power of this formalism is that $\mathcal{A}_n$ is often much simpler than its momentum-space image $\psi_n$.

For $n = 4$ scalars, gauge-fixing $\GL(4)$ and imposing the delta-function constraints reduces~\eqref{eq:Grassint} to a single integration variable~$\tau$:
\begin{equation}\label{eq:tauint}
  \psi_4 = \oint \frac{\text{d}\tau}{2\pi i}\, \mathcal{A}_4(\tau)\,.
\end{equation}
The Grassmannian Mandelstam variables are the three independent $4\times 4$ minors of $C$,
\begin{equation}\label{eq:STUdef}
  S \equiv (\bar 1 \bar 2 1 2)\,,\quad T \equiv (\bar 1 \bar 4 1 4)\,,\quad U \equiv (\bar 1 \bar 3 1 3)\,,
\end{equation}
where ${(I_1 I_2 I_3 I_4) \equiv \epsilon^{abcd}C_{aI_1}C_{bI_2}C_{cI_3}C_{dI_4}}$, with the columns labeled $\bar 1,\cdots,\bar n, 1,\cdots, n$.
Unlike in flat space, the sum of these Grassmannian Mandelstams
\begin{equation}\label{eq:STUsum}
  S + T + U = -2\tau E\,,
\end{equation}
is proportional to the total energy $E{=}k_1{+}k_2{+}k_3{+}k_4$ and vanishes only in the flat-space limit $E\to 0$. 
As functions of $\tau$, each Mandelstam is quadratic in $\tau$
\begin{equation}\label{eq:Stau}
  S(\tau) = \ls(\tau{-}\tau_s)(\tau{-}\tbs)\,,
\end{equation}
where the roots are
\begin{equation}\label{eq:roots}
    \tau_s = \frac{\es}{\ls E}\,,\quad  \tbs = \frac{\ebs}{\ls E}\,,
\end{equation}
with the partial energy products  
\begin{align}\label{eq:partialenergy}
  \es &= (k_1{+}k_2{+}s)(k_3{+}k_4{+}s)\,,\\ \ebs &= (k_1{+}k_2{-}s)(k_3{+}k_4{-}s)\,,
\end{align}
and cyclically for~$t$ and $u$.
The leading coefficients are the spinor helicity bracket products 
\begin{equation}\label{eq:ells}
  \ls \equiv \langle\bar 1\bar 2\rangle\langle\bar 3\bar 4\rangle\,,\quad
  \lt \equiv \langle\bar 1\bar 4\rangle\langle\bar 2\bar 3\rangle\,,\quad
  \lu \equiv \langle\bar 1\bar 3\rangle\langle\bar 4\bar 2\rangle\,.
\end{equation}
In this language, the exchange of a massless spin-$J$ particle in the $s$-channel is given by
\begin{equation}\label{eq:WJ}
  \mathcal{A}_4^{(J,s)} = \frac{S^{J-1}}{(S{+}T{+}U)^J}\, P_J\!\left(\frac{T{-}U}{S}\right),
\end{equation}
where $P_J$ is the Legendre polynomial.
Every such exchange has an order-$J$ pole at $S{+}T{+}U = 0$, corresponding to a total energy singularity $E^{-(2J+1)}$ in momentum space. 
The $\tau$-integral~\eqref{eq:tauint} is defined by the contour prescription of~\cite{Arundine:2026fbr}, which encloses the poles at $\tau = 0, \tbs, \tbt, \bar\tau_u$ in the counter-clockwise direction.
For $J=1,2$ the residue sum reproduces exactly the known momentum-space correlators~\cite{Arundine:2026fbr}, and for $J>2$ we have verified agreement with the corresponding momentum-space results in~\cite{Baumann:2021fxj}.

\section{The Vasiliev Grassmannian}\label{sec:vasiliev}

We now present the central result of this letter.
One cyclic contribution to the Vasiliev Grassmannian correlator is 
\begin{equation}\label{eq:ordered}
  \mathcal{A}_4^{(s,t)} = \frac{U}{ST}\,.
\end{equation}
This expression is to be integrated over the contour~\eqref{eq:tauint} to produce the momentum-space correlator~$\psi_4^{(s,t)}$ in~\eqref{eq:boxmomspace}.
The full crossing-symmetric answer is obtained by summing over the three cyclic contributions in~\eqref{eq:fullcorr}, giving
\begin{equation}\label{eq:fullGrass}
  \mathcal{A}_4^{\rm Vas} = \frac{U}{ST} + \frac{T}{SU} + \frac{S}{TU} = \frac{S^2 + T^2 + U^2}{STU}\,,
\end{equation}
which is our main result~\eqref{eq:mainresult}.

We now show that the contour integral of $U/(ST)$ reproduces the $(s,t)$-contribution to the Vasiliev box~\eqref{eq:boxmomspace}.
The spinor brackets~\eqref{eq:ells} and their unbarred counterparts $\lbs \equiv \langle12\rangle\langle34\rangle$, etc., satisfy the Schouten identity
\begin{equation}\label{eq:schouten}
  \ls + \lt + \lu=0\,,\quad \lbs + \lbt + \lbu=0\,,
\end{equation}
together with the relations
\begin{equation}\label{eq:modulus}
  \ls\lbs = \es\ebs\,,\quad \lt\lbt = \et\ebt\,,\quad \lu\lbu = \eu\ebu\,,
\end{equation}
due to the identity $\langle\bar i\bar j\rangle\langle ij\rangle = (k_i{+}k_j)^2-|\boldsymbol{k}_{i}+\boldsymbol{k}_{j}|^2$. 
Using \eqref{eq:Stau}, ${\cal A}_4^{(s,t)}$ can be expressed as
\begin{equation}\label{eq:split}
  \frac{U(\tau)}{S(\tau)T(\tau)} = \frac{\lu}{\ls\lt}\,\frac{(\tau{-}\tau_u)(\tau{-}\bar\tau_u)}{(\tau{-}\tau_s)(\tau{-}\tbs)(\tau{-}\tau_t)(\tau{-}\tbt)}\,.
\end{equation}
The contour encloses just the poles at $\tbs$ and $\tbt$.
The $\tau$-derivatives at these poles are $S'(\tbs) = -2s$ and $T'(\tbt) = -2t$.
Since $S{+}T{+}U = -2\tau E$, we have $U(\tbs) = -2\tbs E - T(\tbs)$ when $S(\tbs) = 0$, so the residue at~$\tbs$ splits as
\begin{equation}\label{eq:Rsplit}
  R_s = \frac{U(\tbs)}{T(\tbs)(-2s)} = \frac{\tbs E}{s\,T(\tbs)}+\frac{1}{2s}\,,
\end{equation}
and similarly for the residue $R_t$ at $\tbt$. The full residue sum is
\begin{equation}\label{eq:fdef}
  R_s+R_t = f + \frac{1}{2s}+\frac{1}{2t}\,,
\end{equation}
where we defined
\begin{equation}\label{eq:fexpr}
  f \equiv E\!\left[\frac{\tbs}{s\,T(\tbs)}+\frac{\tbt}{t\,S(\tbt)}\right].
\end{equation}
Although $f$ is naively a complicated function of spinor brackets, it reduces to a real scalar, as we show below.

Let us introduce the spinor combinations
\begin{equation}\label{eq:XYdef}
  X \equiv \lt\ebs - \ls\et\,,\quad Y \equiv \ls\ebt - \lt\es\,,
\end{equation}
in terms of which the $\tau$ differences become
\begin{equation}\label{eq:taudiffs}
  \tbs - \tau_t = \frac{X}{\ls\lt E}\,,\quad \tbt - \tau_s = \frac{Y}{\ls\lt E}\,.
\end{equation}
We also define
\begin{equation}\label{eq:eid}
  Q\equiv t\ebs + s\et = t\es + s\ebt \,.
\end{equation}
In terms of these definitions, the function $f$ takes the form
\begin{equation}\label{eq:fclean}
  f = -\frac{Q E^2}{st}\frac{\ls\lt}{XY}\,.
\end{equation}
Expanding $XY$ explicitly, we get
\begin{equation}\label{eq:XYraw}
  XY = \ebs\ebt\,\ls\lt + \es\et\,\ls\lt - \ebt\et\,\ls^2 - \ebs\es\,\lt^2\,.
\end{equation}
The relations~\eqref{eq:modulus} give $\ebt\et\,\ls^2 = \ls\lt\ls\lbt$ and $\ebs\es\,\lt^2 = \ls\lt\lbs\lt$, so~\eqref{eq:XYraw} factorizes as
\begin{equation}\label{eq:XYfactored}
  XY = \ls\lt\bigl(\ebs\ebt + \es\et - \lbt\ls - \lbs\lt\bigr)\,.
\end{equation}
Using the identities~\eqref{eq:schouten} and~\eqref{eq:modulus}, we have
\begin{equation}\label{eq:crossbk}
  \lbs\lt + \lbt\ls = \eu\ebu - \es\ebs - \et\ebt\,,
\end{equation}
which gives a spinor-free expression for the ratio
\begin{equation}\label{eq:XYpartial}
  \frac{XY}{\ls\lt} = \es\et + \ebs\ebt + \es\ebs + \et\ebt - \eu\ebu\,.
\end{equation}
Writing $\es = P_s + sE$ and $\ebs = P_s - sE$ with $P_s \equiv (k_1{+}k_2)(k_3{+}k_4)+s^2$, and using the relation $P_s + P_t + P_u = E^2$ together with the Gram identity $u^2 = \sum k_i^2 - s^2 - t^2$, the right-hand side of~\eqref{eq:XYpartial} evaluates to $2E^2 F_{++}$.
This gives
\begin{equation}\label{eq:XYresult}
  \frac{\ls\lt}{XY} = \frac{1}{2E^2 F_{++}}\,,
\end{equation}
which is a real scalar. 
Substituting~\eqref{eq:XYresult} into~\eqref{eq:fclean} and combining with~\eqref{eq:fdef}, we arrive at
\begin{equation}\label{eq:tauresult}
  \oint\frac{\text{d}\tau}{2\pi i}\frac{U}{ST} = -\frac{Q}{2stF_{++}} + \frac{1}{2s}+\frac{1}{2t}\,.
\end{equation}
Over a common denominator this equals $-\mathcal{N}_4/(2stF_{++})$ and gives
\begin{equation}\label{eq:boxmatch}
  \oint\frac{\text{d}\tau}{2\pi i}\frac{U}{ST} = -\frac{\mathcal{N}_4}{2st F_{++}} = -\frac{1}{2}\psi_4^{(s,t)}\,,
\end{equation}
reproducing the Vasiliev box~\eqref{eq:boxmomspace} (up to a factor of $-1/2$ due to our normalization convention).
Notably, the three spurious singularities $F_{+-}$, $F_{-+}$, $F_{--}$ never appear at any stage, while the Ptolemy singularity $F_{++}$ emerges directly from the integral. 
A similar computation reproduces the two other cyclic contributions $\psi_4^{(t,u)}$ and $\psi_4^{(u,s)}$.

\section{Analytic structure}\label{sec:analytic}

We now comment on the analytic properties of the Vasiliev Grassmannian~\eqref{eq:mainresult}.

\emph{Poles and residues.}~The function $(S^2{+}T^2{+}U^2)/(STU)$ has simple poles at $S = 0$, $T = 0$, and $U = 0$ only.
The residue at each pole takes the universal form
\begin{equation}\label{eq:residues}
  \Res_{S=0}\,\frac{S^2+T^2+U^2}{STU} = \frac{T}{U} + \frac{U}{T}\,,
\end{equation}
together with its cyclic permutations. 
Expanding this in the single exchange basis~\eqref{eq:WJ} with $w=(U{-}T)/(U{+}T)$, one obtains
\begin{equation}\label{eq:UTexpand}
  \frac{T}{U}+\frac{U}{T} = 2\frac{1+w^2}{1-w^2} = \sum_{J=0}^\infty a_J\,\frac{1}{2^J}\binom{2J}{J}\,w^J\,,
\end{equation}
which determines the Grassmannian partial wave coefficients to be
\begin{equation}
    a_0=2\,,\quad a_J = (1+(-1)^J) \frac{2^{J+1}}{\binom{2J}{J}}~~\text{for}~~ J \geq 1\,,
\end{equation}
matching the fact that Vasiliev higher spin gravity has only even spin $J \in 2 \mathbb{Z}_{\geq 0}$ particles in its bulk spectrum. 
This should be contrasted with the individual spin-$J$ exchanges~\eqref{eq:WJ}, each of which carries an order-$J$ pole at $S{+}T{+}U = 0$, or poles at $S{+}T{-}U = 0$ as in the Yang--Mills correlator~\cite{Arundine:2026fbr}.

\emph{Flat-space locus.}~The Vasiliev Grassmannian $\mathcal{A}_4^{\rm Vas}$ can be expressed as 
\begin{equation}\label{eq:deformation}
  \frac{S^2+T^2+U^2}{STU} = -\frac{2}{S} - \frac{2}{T} - \frac{2}{U} + \frac{(S{+}T{+}U)^2}{STU}\,.
\end{equation}
In this form, we see that the departure from flat space is controlled entirely by the cosmological deformation $S{+}T{+}U \neq 0$.
There is, however, no flat-space amplitude to extract: the absence of a total energy pole means there is no residue to interpret as a scattering amplitude.
This is consistent with the expectation that Vasiliev theory has no flat-space $S$-matrix, and should be contrasted with the Grassmannian correlators for Yang--Mills and gravity, both of which have total energy poles, whose residues reproduce the known flat-space amplitudes.

\emph{Uniqueness.}~The most general rational function of $\GL(1)$ weight $-1$ that is crossing-symmetric with at most simple poles at $S,T,U= 0$ is $( \alpha_1\sigma_1^2+\alpha_2\sigma_2)/\sigma_3$, where $\sigma_1\equiv S+T+U$,  $\sigma_2\equiv  ST+TU+US$,  $\sigma_3\equiv STU$. The partial wave coefficients $a_J$, inherited from the free Sp$(N)$ OPE data~\cite{DeLee2026}, fix $\alpha_1 = 1$ and $\alpha_2 = -2$, giving~\eqref{eq:mainresult}.

\section{The Veneziano analogy}\label{sec:veneziano}
The Veneziano amplitude~\cite{Veneziano:1968yb} resums an infinite Regge trajectory into $A^{\text{Ven}}_4{=}\Gamma(-1{-}\alpha's)\Gamma(-1{-}\alpha't)/\Gamma(2{+}\alpha'u)+\text{cyc.}$, where $s,t,u$ here are the usual flat-space Mandelstam variables. The Vasiliev Grassmannian achieves an analogous resummation, collapsing an infinite tower of massless higher-spin exchanges into the rational expression $\mathcal{A}^{\rm{Vas}}_4{=}(S^2{+}T^2{+}U^2)/(STU)$.
In both cases, a small number of physical conditions uniquely determine the answer.
The analogy becomes sharper for the individual cyclic contributions. 
In the field-theory limit $\alpha' \to 0$, the partial Veneziano amplitude reduces to $-u/(st)$, the color-ordered four-point scalar amplitude. 
Remarkably, the corresponding Vasiliev Grassmannian $\mathcal{A}_4^{(s,t)}{=}U/(ST)$ takes exactly this form, with the Grassmannian Mandelstams replacing their flat-space counterparts.

This coincidence is striking and calls for a deeper explanation. Minimal Vasiliev theory is naturally associated with the tensionless limit $\alpha'/L_{\rm (A)dS}^2\to\infty$, where an infinite higher-spin tower becomes massless. Yet, its Grassmannian correlator, when interpreted as a flat-space amplitude, takes the form characteristic of the opposite, field-theory ($\alpha'\to0$) limit! In this sense, the Grassmannian appears to reverse the role of string tension.

Another way to see the connection to the Veneziano amplitude is by fixing three of the $\tau$-roots to $0$, $1$, $\infty$ via a M\"obius map, so that~\eqref{eq:boxmatch} becomes
\begin{equation}\label{eq:KN}
  \psi_4^{(s,t)} = \mathcal{N}\oint_{\Gamma_{01}}\frac{\text{d}x}{2\pi i}\frac{(1 - \alpha x)(1 - \beta x)}{x(1-x)(1-\gamma x)}\,,
\end{equation}
where $x\equiv(\tau-\bar\tau_s)(\bar\tau_t-\tau_t)/[(\tau-\tau_t)(\bar\tau_t-\bar\tau_s)]$, and $\alpha$, $\beta$, $\gamma$ are cross-ratios built out of the zeros of $S,T,U$, and $\mathcal{N}$ accounts for the Jacobian.
The contour $\Gamma_{01}$ encloses the points $x=0$ and $x=1$. 
This is formally the same moduli space $\mathcal{M}_{0,4}$ on which the Koba--Nielsen integral~\cite{Koba:1969rw}
\begin{equation}\label{eq:ven_int}
  A^{\rm Ven}_4 = \int_0^1 \text{d}x\, x^{-\alpha' s - 2}(1{-}x)^{-\alpha' t - 2}\,,
\end{equation}
lives, but with a rational one-form replacing the power-law integrand.
It is natural to ask whether the rational one-form can be deformed into an object with nontrivial monodromy, generating an infinite sequence of Mandelstam poles analogous to a cosmological Regge trajectory. Whether such a deformation exists with the correct analyticity properties remains to be seen.

\section{Outlook}\label{sec:discussion}

The Vasiliev Grassmannian~\eqref{eq:mainresult} is the unique crossing-symmetric rational function with $\GL(1)$ weight $-1$, simple Grassmannian Mandelstam poles, whose residues are fixed by the free Sp$(N)$ higher-spin tower. 
Its simplicity, together with the complete absence of total and partial energy poles present in individual exchanges, suggests that Grassmannian space is the natural language in which the Vasiliev correlator takes its most compact form. 
The fact that a single rational expression captures the exact resummation of an infinite higher-spin tower is striking and points to several natural extensions.

\emph{Other holographic models.}
In AdS, the free O$(N)$ model is dual to Vasiliev theory with $\Delta = 1$ scalar boundary condition. The critical $O(N)$ model, reached by Legendre transform to $\Delta = 2$~\cite{Petkou:2003zz,Giombi:2012ms}, has anomalous dimensions $\gamma_J \sim 1/N$ and modified OPE coefficients~\cite{Giombi:2016hkj, Turiaci:2018nua}. 
It is natural to ask whether the corresponding Grassmannian correlator remains comparably simple, or instead develops new structures such as $\log S$ terms or poles at $S{+}T{+}U = 0$. 
More broadly, the same orthogonal Grassmannian $\OGr(n,2n)$ appears in the ABJM amplitude~\cite{Huang:2013owa}, while the higher-spin limit of ABJ theory~\cite{Chang:2012kt,Hirano:2015yha,Binder:2021cif}, in which a finite Chern-Simons level deforms the free-theory OPE data, provides another controlled setting in which to test these ideas. 
It would be very interesting to understand which features of the Vasiliev Grassmannian are special to the free theory~\cite{Maldacena:2011jn} and which persist more generally~\cite{Maldacena:2012sf}.

\emph{Higher multiplicity.}
The momentum-space scalar five-point correlator is considerably richer, taking the form of a massive pentagon integral with a highly structured numerator~\cite{DeLee2026}. 
Our four-point result suggests that this apparent complexity may again simplify dramatically in Grassmannian space, perhaps into a compact $S_5$-symmetric rational expression. 
At four points, the Grassmannian integral can be viewed as a rational one-form on $\mathcal{M}_{0,4}$. 
At higher points, this simple worldsheet-like picture does not extend directly, and it remains an open question whether the higher-point integrals admit a similar natural geometric description.

\emph{Positive geometry.}~Eq.~\eqref{eq:mainresult} has simple poles only at $S=0$, $T=0$, and $U=0$, and no singularities elsewhere. 
This is the minimal singularity structure one could hope for in a crossing-symmetric function on $\OGr(4,8)$. 
It is therefore tempting to ask whether the Vasiliev Grassmannian admits a positive geometric interpretation~\cite{Arkani-Hamed:2013jha, Arkani-Hamed:2017fdk, He:2023rou}, perhaps as the canonical form of a distinguished positive region.

\emph{Spinning correlators.}~The Grassmannian framework naturally accommodates spinning external states through helicity factors built from Grassmannian minors, such as $(\bar 1\bar 2 3 4)^2$. 
It would be interesting to understand whether the graviton four-point function in Vasiliev theory admits an equally simple Grassmannian representation, with the appropriate helicity structures multiplying the same Mandelstam ratios that appear in the scalar case.
We leave this question for future work.

\smallskip
\emph{Acknowledgments.} 
We thank Nima Arkani-Hamed, Mattia Arundine, Vivek Chakrabhavi, Clifford Cheung, Yu-tin Huang, Javier Huenupi, Austin Joyce, Shani Meynet, Guilherme Pimentel, and Andrew Strominger for useful discussions.
S.D.~is supported by funds provided by the Center for Particle Cosmology at the University
of Pennsylvania. H.L.~is supported in part by the U.S.~Department of Energy under grant DE-SC0013528.

\bibliographystyle{apsrev4-2}
\bibliography{refs} 

\end{document}